\documentclass[12pt,a4paper]{article}

\usepackage[margin=3cm]{geometry}
\usepackage[latin1]{inputenc}
\usepackage[T1]{fontenc}
\usepackage[english]{babel}
\usepackage{amsmath,amsfonts,amssymb}
\usepackage{graphicx}
\usepackage{hyperref}
\usepackage{natbib}

\DeclareMathOperator{\sgn}{sgn}
\def\id{\text{id}}

\title{Signal-wise performance attribution for constrained portfolio optimisation}
\author{Bruno Durin$^{\text{1}}$}
\date{}

\begin{document}

\maketitle
\small
\begin{center}
$^\textrm{1}$~\emph{Capital Fund Management, 23-25 rue de l'Université, Paris, France}\\
\end{center}
\normalsize

\begin{abstract}

Performance analysis, from the external point of view of a client who would only have access to returns and holdings of a fund, evolved towards exact attribution made in the context of portfolio optimisation, which is the internal point of view of a manager controlling all the parameters of this optimisation. Attribution is exact, that-is-to-say no residual ``interaction'' term remains, and various contributions to the optimal portfolio can be identified: predictive signals, constraints, benchmark. However constraints are identified as a separate portfolio and attribution for each signal that are used to predict future returns thus corresponds to unconstrained signal portfolios. We propose a novel attribution method that put predictive signals at the core of attribution and allows to include the effect of constraints in portfolios attributed to every signal. We show how this can be applied to various trading models and portfolio optimisation frameworks and explain what kind of insights such an attribution provides.

\end{abstract}

\section{Introduction}

Performance analysis is at the core of investment process. The pioneering approaches, be they return-based or portfolio-based, took a external stance, aiming at explaining performance with the same data that an investor would have from a fund manager: returns and holdings. These are the long-standing models of performance analysis \citetext{\citealp{Sharpe,Jensen}; see also references in \citealp{Grinold}} and of factor models \citep[][and subsequent works]{FamaFrench} using only time series of returns on the one hand, and the more recent approaches of performance attribution pioneered by \citet{brinson1,brinson2} using both returns and holdings on the other hand.

Return based analysis can be summarised as a regression of fund returns over well chosen and meaningful time series of returns. The coefficients of the regression provide an quantitative assessment of what the manager is doing. As a very simple example, we can check that an index tracker does not have a large cap bias by checking its exposure to the size factor or that a global fund whose prospectus claims balanced exposures to developed markets does not show any oversized exposure to, say, US markets.

Holding based analysis in principle allows for a finer understanding of what the manager is doing. The Brinson et~al. model decomposes the active return\footnote{return over a benchmark} along categories, which originally were sectors, into three components: the allocation part, which corresponds to a strategy that trades benchmark sectors as a whole, the selection part, which corresponds to stock weighting inside a given sector and an interaction part, which simply is the unexplained part. This methodology can be extended to several layers of decision making according to various categories as can be exemplified by industry implementation such as Morningstar's one \citep{Morningstar1,Morningstar2}. However as it is difficult to extend to several categories, a more general framework for performance attribution may be preferred: we regress over portfolio characteristics that can be anything relevant for the analysis, predictive signals as in \citep{GrinoldAttribution} or various factor scores\footnote{for example we could carry a sector, size, value and momentum analysis} for example. In \citep{GrinoldAttribution} these characteristics are translated into portfolios, which allows to express the results in term of risks, correlations and (co-)variances. As it has been noted that Brinson model can be seen as a regression \citep{LuKane}, we shall consider the various holding-based analyses as regressions.

At this point, whatever the level of details at which we perform our analysis, we are basically doing regressions, which have a major drawback: the residual unexplained part may be large. Furthermore, adding many factors to reduce it may lead to in-sample bias and may reduce the explanatory power of the analysis. As shown in the example given in \citep{GrinoldAttribution}, a portfolio built from three signals, a fast one, an intermediate one and a slow one, while taking into account transaction costs, can be explained with a $R^2$ of 87\%. Of course in this case the residual variance is small enough for the analysis to be valuable: it is clear that the portfolio overweights intermediate and slow signals with respect to the ideal, no-cost portfolio, in order to reduce costs as expected. But in the general case the unexplained part can be so large that it is barely possible to conclude anything. Namely this is the case when constraints are imposed on the portfolio and we cannot enlarge them to come to a more amenable situation.

By taking an internal point of view, which means by assuming that not only we have access to returns and holdings of the fund but also to the optimisation procedure used to build the portfolio, we can tackle this problem. In their seminal paper \citep{GrinoldEaston} the authors exactly decompose the performance of a portfolio obtained by constrained mean-variance optimisation into a benchmark part, a signal part and a constraint part. As we shall rephrase it later, the core of the method consists in splitting the optimality equation (KKT\footnote{Karush-Kuhn-Tucker, see for example \citet{Boydcvx} p.~243 and references provided p.~272} condition) into the corresponding terms that can be expressed as what is called in the article characteristics portfolios. Subsequent works \citep{GrinoldIR,SchererXu,StubbsVdbussche,BenderLee} suggested variations and improvements of the method, studying the effect of constraints on key quantities such as information ratio or utility function, addressing alpha  misalignment caused by constraints or taking into account non-linear and/or non-differentiable convex constraints or objective function terms.

Even if these techniques allow exact performance attribution, constraints can only be tackled as separate entities. Let us illustrate this point through a concrete example: a fund manager would like to offer a style shifting value and momentum long-only product to, say, institutional investors. What we mean by style shifting is the fact that the manager has the discretionary power to adjust the relative weight of value and momentum strategies by monitoring their recent performance for example. We assume that the manager knows how to compute value and momentum predictive signals and for a given signal or combination of signals how to build a long/short portfolio through a risk constrained optimisation and a long-only portfolio through a risk and no-short constrained optimisation\footnote{For long-only portfolios the risk constraint would rather be replaced by a tracking error risk constraint and the optimisation be done on portfolio positions relative to benchmark ones, but these are implementation details that are not relevant for the given example.}. How could he build his new product?

One obvious and simple solution would be to add the value long-only portfolio and the momentum long-only portfolio. By monitoring the performance of each portfolios, we would adjust the relative risk attributed to each. But that would be greatly sub-optimal, especially given the fact that value and momentum are anti-correlated\footnote{see for example \citet{ValMom}}, which means that when a position in the value long/short portfolio is long, one expects that the corresponding position in the momentum long/short portfolio is short, but by imposing long-only constraints to both portfolios, we cannot benefit from crossing: if we imposed the constraint on the total portfolio, momentum could take a short position as large as long value position is. If he builds the portfolio this way, by adding predictive signals and running a constrained optimisation to compute the total portfolio, he will run into a different problem: how to attribute performance to value and momentum? The aforementioned techniques allows to attribute a performance to the long-only constraints, but this is likely to be almost as large as the unconstrained value and momentum performances. If the total portfolio is losing money, given the fact that we cannot relax the long-only constraints and assuming that unconstrained value and momentum have both positive performance, which one to cut? In other words, which strategy is most affected by the constraints?

Falling back on regression method is not a solution as it is likely to be useless due to a large residual, be the regressors long/short portfolio performances or long-only ones. Using the exact performance attribution where the constraints are translated into a characteristics portfolio, it's hard to understand the link between the raw performance of a signal and its performance in the constrained portfolio optimisation.

What is usually done in such a case is to use a somewhat ad-hoc scheme to attribute the performance to the signals and to avoid introducing a constraint portfolio whose performance is as large as the one of the signals. It's far easier to make an investment decision based on this attribution. It is not hard to find reasonable schemes of attribution. For example we could attribute a constraint p\&l proportionally to signal absolute size or give a rule to split the total trade into a ``value'' trade and a ``momentum'' trade. But how to advocate them? Which one to choose if two perfectly reasonable schemes lead to contradictory results (in our case study positive and negative performance for the value part of the portfolio for example)?

In this article, we propose an exact signal-wise performance attribution in presence of constraints that allows to overcome the shortcomings of the previous method. Instead of translating the effect of constraints into implied alphas (which is another view of the characteristics portfolios), we show how to translate it into implied costs and risk. To our knowledge, this is the first attempt in this direction. Even though the technique we describe is by no mean the final answer of the problem, it naturally stems from the properties of the problem that we consider and, as an added bonus, yields simple interpretation of the effect that some constraints and terms in the objective function have on the signals.

\section{Yet another look at the Lagrange multipliers}

\subsection{General trading framework}

In this section, we shall review the method of attribution published in \citep{GrinoldEaston}. We shall use the dynamic trading framework of \citet{GP} as a general setting in which to apply the technique. All notations are the same as in their article. Building upon our case study of a long-only style shifting value and momentum fund, we shall consider the following model, which is also given as an example in a more general form (example 2, section V in \citet{GP}):
\begin{equation}
r_{t+1} = B f_t + u_{t+1}\ ,
\end{equation}
where $r$ is the vector of the price changes in excess of the risk-free return of the $N$ equities in our investment universe, $u$ is the unpredictable noise and $f$ is a $2N \times 1$ vector of predictive signals $\begin{pmatrix} v_t & m_t \end{pmatrix}^T$ with $v_t$ (resp. $m_t$) the vector of value (resp. momentum) signals for all stocks. $B$ is a $N \times 2N$ matrix, which for simplicity we will be assume to be $N\times N$ block diagonal, such that we can write for each stock $i$
\begin{equation}
r^i_{t+1} = b_v^i v_t^i + b_m^i m_t^i + u_{t+1}^i\ .
\end{equation}
Redefining $v_t^i$ as $b_v^i v_t^i$ and similarly $m_t^i$, we can simply write the returns as the sum of our two signals and a noise
\begin{equation}
r^i_{t+1} = v_t^i + m_t^i + u_{t+1}^i\ .
\end{equation}
In the general case we will define $K$ vectors $g_k$ for $k=1,\ldots,K$ by $(g_k)^i_t = B_{ik} f^k_t$ in order to write the returns as the sum of $K$ predictive components and a noise
\begin{equation}
r_t = \sum_k g^k_t + u_{t+1} = G_t + u_{t+1} \ .
\end{equation}
The present value of all future expected excess returns penalised for risk and trading costs is maximised by solving a Bellman equation, which is not reproduced here. Taking into account the known solution for the value function and integrating the effect of the dynamics of the predictive signals in their definition, it is not difficult to see that it can be expressed as a quadratic optimisation problem in the trade $\Delta x_t$:
\begin{equation}
\max_{\Delta x_t} -\frac12 \Delta x_t Q \Delta x_t - \Delta x_t P x_{t-1} + \Delta x_t \hat{G}_t\ , \label{master}
\end{equation}
where $\hat{G} = \sum_k \hat{g}_k$ and $\hat{g}_k$ is the predictive components modified by the dynamics of $f$. To clarify this last definition, let us consider the case as in example 2 of section V of \citep{GP} where the mean-reversion speed for a predictive signal $f_k$ is the same for all securities\footnote{Beware that we use indices $i$ for stocks and $k$ for predictive signal/characteristics whereas in \citep{GP}, indices $s$ are used for stocks and $i$ for characteristics. In addition, we implicitly make the assumption that costs are proportional to risk, as done in the article.}. Then
\begin{equation}
\hat{g_k} = \frac1{1+\phi^k a/\gamma} g_k\ .
\end{equation}
Note that we could also consider example 4 of section V (``Today's First Signal is Tomorrow's Second Signal'') by defining $\hat{g_k}$ as the weighted sum of future signals $g_k$ as shown in equation (30) of their article.

We do not give the general expressions for $Q$ and $P$ as we do not need them in what follows. However to illustrate what they are, let us consider what is called the Static Model in example 3, section V of \citet{GP}: the same optimisation problem has to be solved, with matrices $Q$ and $P$ whose expressions are given below:
\begin{equation}
Q = \Lambda + \gamma \Sigma = (\lambda +\gamma) \Sigma\ , \qquad P = \gamma \Sigma\ .
\end{equation}
$\Lambda$ is the matrix of quadratic costs, $\Sigma$ is the quadratic risk matrix, $\gamma$ the risk aversion. Second equation for $Q$ corresponds to the assumption that risk and costs are proportional, which simplifies the results a lot (we introduce $\lambda$ the cost coefficient such that $\Lambda = \lambda \Sigma$).

Through this short review of \citep{GP}, we emphasised that the quadratic optimisation problem \eqref{master} is the common problem to solve for various models of trading.
\begin{itemize}
\item Quadratic risk: this is the risk measure proposed by Markowitz and is widely used.
\item Quadratic costs: impact is generally modelled as a square-root function of the trade, which corresponds to cost terms with power 3/2 \citep{Almgrenq12,Engleq12,Abdobalq12,Kisselq12,Moroq12,Tothq12th}. However, as mentioned in \citep{GP}, calibration of a quadratic cost model has also been done.
\item Possibly persistent price impact costs: for the purpose of what follows, let us simply note that the price distortion $D_{t}$ induces a cross term $D_{t-1}\Delta x_t$ similar to the one for position $x_{t-1}$ and in all what follows $D_{t-1}$ should be treated in the same way as $x_{t-1}$. We shall give some indications for this below.
\item Dynamic or static model (taking into account the future expected predictive signals or not) by using either $\hat{G}$ or $G$ and the corresponding expressions for $Q$ and $P$.
\end{itemize}
As we build our attribution technique upon the properties of this optimisation problem, it is only relevant for trading models and portfolio constructions that are described by the optimisation problem. However we cover most of what is suggested in the literature and used by the practitioners.

In all what follows we do not distinguish between various models. Namely $\hat{G}$ will not be distinguished from $G$.

\subsection{Linearity}
\label{linearityexplained}

The key property of the optimality equation of \eqref{master}
\begin{equation}
Q \Delta x_t + P x_{t-1} = G_t \label{optimality}
\end{equation}
is linearity. $Q \Delta + P$ is a linear operator over the time series $\{x_t\}$ of all positions\footnote{In a continuous time setting, we would have a linear first order differential equation with unknown function $x(t)$.}.
Linearity allows us to write the solution $x_t$ as
\begin{equation}
x_t = \sum_k x^k_t \label{sumpos}
\end{equation}
where $x^k_t$ is the solution of the same equation with the source term $G_t$ replaced by component $g^k_t$
\begin{equation}
Q \Delta x^k_t + P x^k_{t-1} = g^k_t \ .
\end{equation}
Computing the solution is straightforward through a iterative process. Let us start at $t=0$ with zero total position $x_0 = 0$ and zero position on all components $x^k_0 =0$. We compute each trade component:
\begin{equation}
\Delta x^k_t = Q^{-1} (g^k_t - P x^k_{t-1})
\end{equation}
and update each position components
\begin{equation}
x^k_t = x^k_{t-1} + \Delta x^k_t\ .
\end{equation}
Should we add a persistent impact, we would assume that at $t=0$, total price distortion is $D_0 = 0$ and its components $D^k_0 = 0$ and we would update each components with the evolution equation for the price distortion\footnote{Evolution equation of $D$ encodes a linear exponential kernel operator on trades $\{\Delta x_u\}_{u\leqslant t}$ which would replace operator $Q\Delta$. When linearity property is applied to the resulting optimality equation, it translates into splitting $D$ into $K$ components $D^k$.}: $D^k_{t+1} = (I-R)(D^k_t + C \Delta x^k_t)$. The meaning of the notations is given in \citep{GP} and is not crucial for what we explain.
We update trades and positions for every signal, whose sum is exactly the total optimal trade and total optimal position \eqref{sumpos}.

We remind that such an attribution has already been described in \citep{GrinoldEaston}. We insist on the fact that it merely is a consequence of the linearity of our optimality equation and could be applied to any optimisation problem, convex or not, whose optimality equation is linear.

Through linearity, we get a direct unambiguous and exact attribution to the various components, which are added up to predict the future returns. From the position time series $\{x^k_t\}$ it is straightforward to compute a p\&l and to attribute risk and costs following \citep{Litterman} or \citep{BruderRoncalli}:
\begin{equation}
R = x_t \Sigma x_t = \sum_k (x^k_t \Sigma x_t) = \sum_k R_k
\end{equation}
and
\begin{equation}
C = \frac12 \Delta x_t \Lambda \Delta x_t = \sum_k \left(\frac12 \Delta x^k_t \Lambda \Delta x_t\right) = \sum_k C_k\ .
\end{equation}

\subsection{Constraints and additional terms in the objective function}
\label{cstr_costs}

The linearity property of optimality equation indeed allows us to attribute trades and positions to various constraints as the authors of \citep{GrinoldEaston} did in introducing their characteristic portfolios. Not all constraints fit in this framework, but most of those used in portfolio optimisation do. If we let aside combinatorial constraints (number of trades, round-lots, etc.) and non convex constraints (minimum trade size for example), the usual suspects are:
\begin{itemize}
\item minimum and maximum trade, minimum and maximum position
\begin{equation}
m_i \leqslant \Delta x_t \leqslant M_i \qquad m_i \leqslant x_{t-1} + \Delta x_t \leqslant M_i
\end{equation}
where usually $m_i = -M_i$ (that is to say, only trade or position size is constrained),
\item minimum and maximum exposure of the portfolio
\begin{equation}
m \leqslant (x_{t-1} + \Delta x_t)\cdot v \leqslant M
\end{equation}
where $v$ is a vector encoding the exposure. As an example, for a sector exposure, $v$ would be 1 for stocks belonging to the given sector and 0 elsewhere. Another example can be found in the usual formulation of the minimum-variance problem where the constraint that net exposure should be 1 is imposed: $v$ would be a vector of 1 and we should set $m = M = 1$ (and prediction $G=0$). Last example, imposing market neutrality would lead us to choose $v$ as the vector of stock betas. Exposure constraints are pervasive in portfolio optimisation problems.
\end{itemize}
All these constraints can be written as $f(\Delta x_t) \leqslant M$ where $f$ is a linear function of $\Delta x_t$: let $v$ be the vector such that $f(\Delta x_t) = v\cdot \Delta x_t$. We shall label constraints with index $c$ and consider the collections of constraint vectors $v_c$ and bounds $M_c$.

We now turn back to our optimisation problem \eqref{master}, which we add constraints to. This is straightforward for the static model. For the dynamic model considered in \citep{GP}, constraining only the next-step optimisation of $\Delta x_t$ while using the unconstrained solution of the Bellman equation for the value function is a standard approximation \citep[see as examples][]{sznaier, skafboyd}. Constrained dynamic programming problems are notoriously difficult to solve\footnote{but see \citep{Bemporad} in which the authors show that control policy for linear constraints are linear functions $f(y) = Ay+b$ of the state $(G_t, x_{t-1})$, which let us think that what is described in this article is also applicable when the value function is not approximated.} and will not be considered here. Introducing Lagrangian multipliers $\lambda^c$ as done in \citep{GrinoldEaston}, optimality equation \eqref{optimality} becomes
\begin{equation}
Q \Delta x_t + P x_{t-1} = G_t + \sum_c \lambda^c_t \ .
\end{equation}
$\lambda^c_t$ are additional sources for which we can define trades $\Delta x^c_t$ and iteratively build positions $x^c_t$ as explained before for the $K$ predictive signals $g^k$. It is straightforward to attribute performance, risk and costs to the portfolios associated to each constraint or group of constraints.

Applying this method to our long-only style shifting value and momentum fund, the optimality equation reads
\begin{equation}
Q \Delta x_t + P x_{t-1} = v_t + m_t + \lambda_t
\end{equation}
where $\lambda_t$ is the vector of all Lagrange multipliers associated to long-only constraints (we choose to consider long-only constraints as a whole as far as attribution is concerned). Performance is attributed as follows: the performance of (unconstrained) value portfolio, the performance of (unconstrained) momentum portfolio and the performance of the long-only constraint portfolio. As we noted above, it may be quite difficult to understand which signal is performing best in presence of the constraint. Overall performance is likely to be far from the one of value or momentum portfolios, the performance attributed to long-only constraints contributing as a large negative bias to the unconstrained portfolio performances. It is hard to tell from this attribution which predictive signal is most hampered by the constraint.

More generally, it is not always meaningful to distinguish a constraint. For example if we considered a quadratic risk constraint instead of a fixed risk aversion ($\gamma$ is then a Lagrange multiplier and is dependent on time $t$), we would get on the one hand the portfolio associated to the risk constraint and on the other hand completely unconstrained predictive signal portfolio whose risk is unbounded. The performance attribution is very likely to look like the sum of two random walks! In this case, it is straightforward (at least in the static model) to include the effect of the risk constraint in signal portfolios as the optimality equation is the same for a time-dependent $\gamma_t$.

Before turning to our suggested solution to this problem, let us show how other terms or constraints can be included in this framework. Non differentiable constraints / terms are dealt with in a different manner from the one in \citep{StubbsVdbussche} where sub-gradients are used and introduce some added complications. We shall consider two types of terms.
\begin{itemize}
\item Non quadratic costs in the objective function:
\begin{itemize}
\item $L^1$ costs (usually associated to bid-ask spread) or turnover constraint: a term $-\lambda_0 \sum_i \lvert \Delta x^i_t \rvert$ where $\lambda_0$ is the cost normalisation or the Lagrangian multiplier associated to the constraint,
\item square-root impact costs with a term $-\Delta x_t \Lambda_{1/2} (\Delta x_t)^{1/2}$, where the power is a signed power ($x^{1/2} = \sgn(x) \sqrt{\lvert x \rvert}$). Note that in the literature $\Lambda_{1/2}$ is usually taken to be proportional to identity and the term reads $-\lambda_{1/2} \lvert \Delta x_t \rvert^{3/2}$.
\end{itemize}
\item Financing costs or leverage constraint ($L^1$ costs for position): a term $-\lambda_l \sum_i \lvert x^i_{t-1} + \Delta x^i_t \rvert$ where $\lambda_l$ is the half-spread between long and short financing for a strictly market neutral portfolio\footnote{In a more general setting, financing of the long positions and financing of the short positions lead to two distinct terms, the first being function of $(x^i_{t-1} + \Delta x^i_t)_+ = \max (x^i_{t-1} + \Delta x^i_t,0) = \frac12 \left(x^i_{t-1} + \Delta x^i_t + \lvert x^i_{t-1} + \Delta x^i_t \rvert\right)$ and the second being function of $(x^i_{t-1} + \Delta x^i_t)_- = \min (x^i_{t-1} + \Delta x^i_t,0) = \frac12 \left(x^i_{t-1} + \Delta x^i_t - \lvert x^i_{t-1} + \Delta x^i_t \rvert\right)$. An optimisation problem with such terms can also be turned into a more simple one as described after.} or the Lagrangian multiplier associated to the constraint.
\end{itemize}
Note that $L^1$-terms may appear even in absence of any financial costs in the problem: they can be introduced as regularisation terms in the context of a lasso regression \citep{Tibshirani}. For an application to portfolio optimisation, see \citep{DeMiguel}, which focuses on minimum variance portfolio, and the more recent \citep{Bruder}. 

Square-root impact costs are considered in the appendix. We shall here focus on the $L^1$ terms / constraints. When such terms are added, the optimisation problem \eqref{master} can be turned into a more simple one by introducing auxiliary variables. This is a standard procedure (see \citet{Boydcvx} section 6.1.1 example ``Sum of absolute residuals approximation'' or \citep{cvxL1costs} for a variant that is more suitable for constraints only on short or long positions). The optimisation problem:
\begin{equation}
\max_{\Delta x_t} -\lambda_0 \sum_i \lvert \Delta x^i_t \rvert -\lambda_l \sum_i \lvert x^i_{t-1} + \Delta x^i_t \rvert -\frac12 \Delta x_t Q \Delta x_t - \Delta x_t P x_{t-1} + \Delta x_t G_t \label{masterL1}
\end{equation}
is equivalent to the following one
\begin{equation}
\begin{gathered}
\max_{\Delta x_t, s, u} -\lambda_0 \sum_i s_i -\lambda_l \sum_i u_i -\frac12 \Delta x_t Q \Delta x_t - \Delta x_t P x_{t-1} + \Delta x_t G_t \\
-s_i \leqslant \Delta x^i_t \leqslant s_i \\
-u_i \leqslant x^i_t + \Delta x^i_t \leqslant u_i \ .
\end{gathered} \label{L1cstr}
\end{equation}
The additional constraints are also linear in $\Delta x^i_t$ which means that the Lagrangian multipliers $\xi_c$ associated to them will appear as additional sources of the linear optimality equation:
\begin{equation}
Q \Delta x_t + P x_{t-1} = G_t + \sum_c \xi^c_t
\end{equation}
where $c$ indexes the type of constraint such that $\xi^c_t$ is the vector of all Lagrangian multipliers associated the constraints of the given type applied to each stock.

Cost attribution can be generalised in the following way:
\begin{equation}
-\lambda_0 \lvert \Delta x_t \rvert = - \sum_k \left[\lambda_0 \Delta x^k_t  \sgn (\Delta x_t)\right]
\end{equation}
and
\begin{equation}
-\lambda_l \lvert x_{t-1} + \Delta x_t \rvert = - \sum_k \left[\lambda_l (x^k_{t-1} + \Delta x^k_t) \sgn (x_{t-1} + \Delta x_t)\right]
\end{equation}
where we define $\sgn(y) = 0$ for $y=0$.

With standard performance attribution that associates a portfolio to each constraint type, it is hard to get a clear interpretation of these terms. As there is a portfolio associated to spread costs for example, the decomposition of costs in the p\&l attributes quadratic costs and spread costs to this portfolio as it does to the others (portfolios associated to signal components and to other constraints). What is the meaning of quadratic costs for the portfolio associated to spread costs? What are these spread costs associated to spread cost portfolio? Has this question even got a meaning?

What is the interpretation of the risk associated to spread cost portfolio? For this question we could make an educated guess: we could understand it as a risk reduction associated with the costs that prevents us from making trades as big as we would have done if these additional costs were not present.

Nevertheless, the asymmetry that such a decomposition introduces between quadratic costs and spread costs is hard to justify. It would be more natural to directly get the combined effect of quadratic and spread costs on a given signal.

\section{Signal-wise attribution of constraints}

\subsection{Effective quadratic costs and effective quadratic risk}

It is now clear that we would like an exact attribution that does not introduce additional portfolios for constraints or terms that are converted into constraints. We shall show that we are able to express all the constraints and additional terms that we listed in the previous section as effective quadratic costs and effective quadratic risk. The optimality equation will have the form
\begin{equation}
\bar{Q}_t \Delta x_t + \bar{P}_t x_{t-1} = \sum_k g^k_t \label{swa}
\end{equation}
where the only source terms are the predictive signals. To our knowledge, the technique we shall describe is original, but the idea of considering the effect of constraints as a deformation of the quadratic risk has already been presented. In \citep{JMa,Roncallishrink}, the authors show how minimum and maximum position constraints can be seen as a shrinkage of the covariance matrix used in the optimisation problem. The key element that allows them to do this is a constraint on the net exposure of the portfolio: $\mathbf{1}\cdot x_t = 1$. Building upon this idea, the author of \citep{deBoer} generalises the work of \citep{JMa} by showing how constraints imply a ``shrinkage estimate'' of the mean and covariance of returns. His work allows to consider more general constraints but as it uses the same mathematical framework it suffers from the same shortcomings.

We present a generalisation of these results through a new method. First of all we take into account transaction costs and show how some very widespread constraints translate into effective quadratic costs. Furthermore in our framework there is no need for any specific constraint (such as $\mathbf{1}\cdot x_t = 1$) that in the previous works is key to build the effective risk matrix. Last but not least we do not affect the estimation of the mean of returns: there is no such notion as implied alpha or shrunk alpha, as to our mind managers do not use constraints to improve return estimations, but rather to control for the effect of a bad risk estimation (namely inverting a badly estimated risk matrix as done in Markowitz optimisation problem introduces a lot of noise). By taking full advantage of the mathematical properties of the optimisation problem, we developed an original method that is less dependent on some specific characteristics of the problem and that allows for a more direct and natural attribution\footnote{As it will be obvious in what follows, it is also straightforward to see that the effective risk matrix is positive-definite and to come to the shrinkage interpretation of the original quadratic risk matrix towards a diagonal risk matrix.}.

All constraints and terms we considered in last section can be put under a linear constraint form,
\begin{equation}
v \cdot \Delta x_t \leqslant M
\end{equation}
at the expense of introducing auxiliary variables in some cases. We shall explicitly distinguish between constraints on trade and constraints on position
\begin{equation}
v \cdot (x_{t-1} + \Delta x_t) \leqslant M
\end{equation}
We shall describe the technique on position constraints. Adaptation to trade constraints is straightforward.

Generally speaking, constraints will go in pairs:
\begin{equation}
m \leqslant v \cdot (x_{t-1} + \Delta x_t) \leqslant M
\end{equation}
One of the following constraint
\begin{equation}
\left(v \cdot (x_{t-1} + \Delta x_t)\right)^2 \leqslant m^2 \text{ or }  \left(v \cdot (x_{t-1} + \Delta x_t)\right)^2 \leqslant M^2
\end{equation}
is equivalent the previous one: when upper bound and lower bound are defined, only one of the bounds is active at the same time.
Now let us introduce Lagrangian multipliers $\eta$ for the equivalent constraint. The KKT conditions for optimality consist in finding the critical point of augmented objective function $\cal F$
\begin{equation}
{\cal F} = -\frac12 \Delta x_t Q \Delta x_t - \Delta x_t P x_{t-1} + \Delta x_t G_t - \eta\, (x_{t-1} + \Delta x_t)\, v \otimes v\, (x_{t-1} + \Delta x_t)
\end{equation}

In the static model where $Q$ is the sum of quadratic costs $\Lambda$ and penalised quadratic risk $\gamma \Sigma$ and $P$ is the penalised quadratic risk, it is obvious that under such a form the position constraint is equivalent to an additional quadratic risk $2\eta/\gamma\, v \otimes v$. Effective quadratic risk is thus $\Sigma + 2\eta/\gamma\, v \otimes v$. Interpretation is the following: if a constraint is violated, we add a factor to the risk model, whose strength we tune to reduce the exposure to the authorised level. If the constraint is a simple minimum or maximum position $m \leqslant x_t \leqslant M$, the effective risk that it introduces simply is an ad-hoc idiosyncratic risk for the stock. If we have only constraints on positions, it is easy to see that we effectively perform a shrinkage towards a diagonal risk matrix. We get a result that yields the same interpretation as in \citep{JMa}. Furthermore, as we showed that adding constraints is equivalent to adding factors in the risk model, we can shed new light on the factor alignment problems: reversing the process, we could try to understand the solutions advocated in \citep{LeeStefek,BLS,SaxenaStubbs1,SaxenaStubbs2,CSS} in terms of constraints on the non-aligned part of the optimal portfolio.

Squaring trade constraints will lead to effective quadratic costs, whose interpretation is even more straightforward.

In the dynamic trading framework, position (resp. trade) constraints also lead to effective quadratic risk (resp. costs) but the expression of the effective quadratic risk (resp. costs) as a function of the original quadratic risk (resp. costs) and the penalty term introduced by the constraint is not simple to establish in the general case. This could be the purpose of a future work.

In the presence of all constraints, the optimality equation reads
\begin{equation}
(Q + \sum_c \mu^c_t A_c + \sum_{c'} \eta^{c'}_t A_{c'}) \Delta x_t + (P + \sum_{c'} \eta^{c'}_t A_{c'}) x_{t-1} = \sum_k g^k_t
\end{equation}
which is the form \eqref{swa} that we announced before, provided one defines $\bar{Q}_t$ and $\bar{P}_t$ as
\begin{equation}
\begin{aligned}
\bar{Q}_t &= Q + \sum_c \mu^c_t A_c + \sum_{c'} \eta^{c'}_t A_{c'} \\
\bar{P}_t &= P + \sum_{c'} \eta^{c'}_t A_{c'}
\end{aligned}
\end{equation}
As for notations, $c$ indexes trade constraints, $c'$ indexes position constraints. $\mu_c$ are the Lagrangian multipliers associated to squared trade constraints, whereas $\eta_{c'}$ are those associated to squared position constraints. Matrices $A_c$ are defined as $2\, v_c \otimes v_c$. For example in the simple case of a minimum or maximum trade or position constraint on stock $i$, $A_c$ is a matrix whose diagonal element $i,i$ is equal to 1 and whose all other elements are 0.

\subsection{Attribution}

We established a linear equation \eqref{swa} whose only source terms are the predictive signals. As explained in subsection \ref{linearityexplained}, we can directly attribute trades and positions to the $K$ signals. We  obtain $K$ portfolios, one for each signal, which includes the effect of constraints and cost terms. Risk and costs attribution is now only done over the predictive signal portfolios, which avoids some of the inconsistencies we mentioned earlier.

Let us explicitly work out our running example of the long-only style-shifting value and momentum fund. Risk is effectively increased on all positions that would be short\footnote{Let us remind that positions are over the benchmark so long-only constraints are in fact lower bounds that are equal to minus the benchmark positions.} in absence of the long-only constraints until they are equal to 0. Let us call $\cal C$ the set of stocks for which long-only constraint is active and define $\rho_t$ a diagonal matrix whose diagonal element $i,i$ is equal to $2\eta^i_t$ if $i \in {\cal C}$ else 0. Optimality equation \eqref{swa} reads
\begin{equation}
(Q + \rho_t) \Delta x_t  + (P + \rho_t) x_{t-1} = v_t + m_t
\end{equation}
Let us assume for simplicity that costs are proportional to risk, that risk is diagonal $\Sigma = \sigma^2$ and that we are using the static model. In this case, in the absence of constraints:
\begin{equation}
\Delta x_t = \frac{\gamma\sigma^2}{\gamma\sigma^2 + \lambda}(x^0_t - x_{t-1})
\end{equation}
where $x^0_t$ is the Markowitz solution $G_t/(\gamma \sigma^2)$. In the presence of constraints, our attribution technique yields two trades for value and momentum
\begin{equation}
\begin{aligned}
\Delta x_{v, t} &= \frac{\gamma\sigma^2 + \rho_t}{\gamma\sigma^2 + \rho_t + \lambda}(x^0_{v,t} - x_{v,t-1}) \\
\Delta x_{m, t} &= \frac{\gamma\sigma^2 + \rho_t}{\gamma\sigma^2 + \rho_t + \lambda}(x^0_{m,t} - x_{m,t-1})
\end{aligned}
\end{equation}
In the expression for $x^0_{v,t}$ and $x^0_{m,t}$, $\gamma\sigma^2$ is replaced by $\gamma\sigma^2 + \rho_t$. If a stock is constrained, its aim position\footnote{as named in \citep{GP}} $x^0$ is reduced in absolute value and the trade will tend to get closer to this corrected aim so that constraint is fulfilled. The impact of the constraint on value and momentum trade depends not only on signal strength or the aim position but also on the current position reached by previous trades.

Returning to the general interpretation as effective risk, for constrained stocks value and momentum signals are run with an effectively higher risk aversion, which means that even if the trade of one of the two signals is in the right direction (long for a minimum position constraint), it is affected by the constraints. This is in contrast to a more naive ad-hoc attribution that for example would consist in computing unconstrained signal trades and in cutting only the one going short. But that is exactly the difference between seeing constraints as a shift in predicted returns\footnote{Shifted alpha is called implied alpha in the literature.} and seeing (position) constraints as a shrinkage of risk estimation.

This direct and exact signal-wise attribution allows us to track performance, risk and costs for each predictive signals. We are able to make decisions such as signal weighting even in presence of strong constraints. We are also able to compute a transfer coefficient for each predictive signal and either drop signals whose coefficient is too low, meaning that constraints are too strong for them to deliver their performance in presence of other signals, or relieve some constraints to let signals ``breathe'' better.
As an example a manager could adjust constraints so that a factor or style timing strategy really has a value-added or our manager running a value and momentum long-only fund could realise that constraints make the addition of, say, a growth strategy useless.

\subsection{How to compute effective costs and risk?}

Let us come back to a more practical point of view. How are the Lagrangian multipliers of squared constraints, which we shall call attribution multipliers in contrast with the Lagrangian multipliers of original constraints, computed?

Firstly let us note that we could in principle design an ad-hoc penalty-like optimisation algorithm that would work as follows:
\begin{itemize}
\item compute unconstrained optimal trades
\item for all trades (resp. positions) that violate a constraint, add a penalty term\\ $\eta_c \Delta x_t A_c \Delta x_t$ (resp.  $\eta_c (x_{t-1} + \Delta x_t ) A_c (x_{t-1} + \Delta x_t)$)
\item compute the corresponding optimal trades
\item update $\eta_c$ until constraints are fulfilled
\end{itemize}
Such a naive algorithm provides no convergence bound. As the original constrained problem is convex in most of the cases, it would be a pity that we could not use the power of the numerous algorithms that exist to solve it. One of the specific cases where the ad-hoc algorithm might be interesting is the minimum trade size constraint, which is not convex. This might be handled by allowing $\eta_c$ to be negative, that is to say to allow for negative effective costs. Indeed if a non zero trade is rounded up to the minimum trade size, this amounts to compute an unconstrained trade with smaller quadratic costs: hence negative costs have been added to the original quadratic costs. We shall not pursue this idea here and shall only consider convex constraints.

From this point, we shall assume that the constrained optimisation problem has been solved by an algorithm that provides both optimal trades and Lagrange multipliers $\lambda^c$ of the original constraints. Lagrange multipliers encode the marginal variation of the objective function at the optimum for a marginal variation of constraint bound:
\begin{equation}
\lambda^c = \epsilon \frac{\partial {\cal F}^\star}{\partial M_c}
\end{equation}
where $\epsilon$ is a sign, equal to $+1$ if the constraint is an upper bound\footnote{The given signs correspond to a maximisation problem.} and equal to $-1$ if the constraint is a lower bound. For attribution multipliers, we have
\begin{equation}
\eta^c = \frac{\partial {\cal F}^\star}{\partial (M_c^2)}
\end{equation}
where sign is always $+1$ as $M_c^2$ is always an upper bound and we have the same ${\cal F}^\star$ as it is an equivalent optimisation problem leading to the same solution. As $\partial/\partial (M_c^2) = 1/(2 M_c) \partial/\partial M_c$, we get the following relationship between both multipliers
\begin{equation}
\eta^c = \epsilon \frac1{2 M_c}\, \lambda^c \label{attrmult}\ .
\end{equation}
It can be checked that under mild assumptions if constraints are correctly split into trade and position constraints, $\eta_c$ is positive. Or we can pragmatically turn this around for non straightforward cases, choose to see the constraint as a trade (an effective cost) or a position (an effective risk) constraint so that the corresponding multiplier is positive. For this to be always possible, the constraint must reduce either the trade or the position, which is the case for all trade and position constraints whose admissible space contains 0.
The relationship \eqref{attrmult} can also be understood in a way reminiscent of what is done in \citep{JMa}\footnote{which could roughly be summarised as using the constraint $\mathbf{1}\cdot x = 1$ to introduce a dependence on $x$ in the optimality equation, such dependence being interpreted as coming from an effective quadratic risk}: we introduce in the original optimality conditions an explicit dependence on trade for a trade constraint (and similarly for a position constraint) by using the following substitution
\begin{equation}
1 = \frac{2 v\cdot \Delta x_t }{2 M_c} \ ,
\end{equation}
which is true whenever the constraint is saturated. For example the $\lambda_c v_c$ term that would appear in the original optimality equation for an upper bound constraint on trade can be turned into $2\,\eta^c\, v \otimes v\, \Delta x_t$. If the constraint is saturated the previous equation holds, otherwise  $\lambda^c$ and $\eta^c$ are zero.

From the solution provided by the algorithm, which we assumed to provide Lagrangian multipliers, it is straightforward to compute attribution multipliers, to build the optimality equation \eqref{swa} and to perform the signal-wise attribution.

The relationship \eqref{attrmult} highlights a corner case that we overlooked. How to deal with the case when the bound is zero? $\eta_c$ is infinite in this case. But this is not a problem, neither from a mathematical point of view nor from an interpretation point of view. Let us start with the latter. If a constraint that sets a trade to 0 is active, it indeed corresponds to infinite quadratic costs. Similarly to force a position to 0, quadratic risk must be infinite (or risk aversion must be infinite). From the mathematical point of view, we shall show that the limit is perfectly regular. This means that we should consider $\eta_c$ as elements of a projective space and find a way to deal with infinite values in the optimality equation \eqref{swa}\footnote{One way to do this is to set the corresponding $\eta_c$ to a moderately large value so that they are large in front of the others $\eta_c$ while preventing the linear system from becoming ill-conditioned. Another more complicated way would be to explicitly deal with those constraints, which are often equality constraints (zero trade or zero position).}.

Now let us turn to the mathematical point of view. Without loss of generality, let us consider the case of a trade constraint: $v \cdot \Delta \pi \leqslant M$. Optimality equation for signal-wise attribution \eqref{swa} reads
\begin{equation}
(Q + 2\eta\, v\otimes v) \Delta x_t = G_t - P x_{t-1} \ .
\end{equation}
Sherman-Morrison formula for the inverse of a rank-1 update of an invertible matrix leads to
\begin{equation}
\Delta x_t = \left(Q^{-1} - \frac{2\eta\, Q^{-1}\, v\otimes v\, Q^{-1}}{1+ 2\eta\, vQ^{-1}v}\right) (G_t - P x_{t-1}) \ .
\end{equation}
Defining $\alpha$ as
\begin{equation}
\alpha = \frac{2\eta\, vQ^{-1}v}{1+ 2\eta\, vQ^{-1}v}\ ,
\end{equation}
the result can be written as
\begin{equation}
\Delta x_t = (1-\alpha) Q^{-1} (G_t - P x_{t-1}) + \alpha \left(Q^{-1} - \frac{Q^{-1}\, v\otimes v\, Q^{-1}}{vQ^{-1}v}\right) (G_t - P x_{t-1}) \ .
\end{equation}

This result yields a geometrical interpretation of the effect of the squared constraint. The trade is a weighted sum of the unconstrained trade and of the result of the projection of the unconstrained trade over the subspace orthogonal to $v$ along the direction $Q^{-1}v$, which takes into account risk and costs (this is not an orthogonal projection). As the bound $M$ goes to 0 and $\eta \to \infty$, we see that the weight $\alpha$ of the projected trades goes to 1 whereas that of the unconstrained trades goes to 0. The limit is well defined and is easily interpreted as simply being the projection on the subspace orthogonal to $v$, which is consistent with the constraint: $v \cdot \Delta x = 0$. In this limit we could have directly guessed that the result should be a projection, but the direction along which to do the projection is not trivial.

For a position constraint, a similar computation can be done. The optimality equation reads
\begin{equation}
(Q + 2\eta\, v\otimes v) \Delta x_t = G_t - (P + 2\eta\, v\otimes v) x_{t-1} 
\end{equation}
and the solution can be written as
\begin{equation}
\begin{split}
\Delta x_t = & (1-\alpha) Q^{-1} (G_t - P x_{t-1}) + \alpha \left(Q^{-1} - \frac{Q^{-1}\, v\otimes v\, Q^{-1}}{vQ^{-1}v}\right) (G_t - P x_{t-1}) \\
 & - \alpha \frac{Q^{-1}\, v\otimes v}{vQ^{-1}v} x_{t-1}\ .
\end{split}
\end{equation}
The trade is a weighted sum of three terms. The first two are the same as for the trade constraint: unconstrained and projected unconstrained trade. The third one is the trade that should be done to project initial position $x_{t-1}$ on the subspace orthogonal to $v$ along direction $Q^{-1} v$. Once again, this is consistent with the constraint $v \cdot (x_{t-1} + \Delta x_t) = 0$ in the limit where the bound goes to 0 ($\alpha \to 1$).

\subsection{Interpretation of the performance attribution of $L^1$ constraints}

In this subsection, we shall give a detailed account of the treatment of $L^1$ constraints or terms (spread costs / turnover constraint, financing cost / leverage constraint) and shall give an interpretation of the effect of such constraints on the predictive signals.

Let us begin with $L^1$ trade terms. As explained in subsection \ref{cstr_costs}, the term $-\lambda_0 \sum_i \lvert \Delta x^i_t \rvert$ in the objective function is turned into linear term and constraints by introducing auxiliary variables $s_i$ (see \eqref{L1cstr})
\begin{equation}
-\lambda_0 \sum_i s_i \quad \text{with constraints} \quad -s_i \leqslant \Delta x^i \leqslant s_i
\end{equation}
where we dropped time indices.
The squared constraints add the following term 
\begin{equation}
-\eta_i \left[(\Delta x^i)^2 - s_i^2 \right]
\end{equation}
in the augmented objective function whose critical point is the optimum of the constrained problem. This critical point is given for $\Delta x^i$ by the optimality equation \eqref{swa} and for auxiliary variable $s_i$ by
\begin{equation}
-\lambda_0 + 2\eta_i s_i = 0 \ . \label{eq_on_s}
\end{equation}
This is the same relation \eqref{attrmult} for $\eta$ as for other constraints: bound $M_c$ is replaced by bound $s_i$:
\begin{equation}
\eta_i = \frac1{2 s_i} \lambda_0 \ .
\end{equation}
This relation holds, be $\lambda_0$ a fixed spread cost or the Lagrangian multiplier of a turnover constraint. As $s_i = \lvert \Delta x_i \rvert$, it is as easy to compute $\eta_i$ from the solution given by a solver as for the other constraints. If the trade is zero, $\eta_i \to \infty$, spread costs acts as infinite effective quadratic costs.

As spread costs can be seen as a threshold that the total predictive signal $G$ must overcome, one might wonder how this threshold behaviour appears with effective quadratic costs. For simplicity, we shall only consider one stock in the static model. The optimisation problem reads
\begin{equation}
\max_{\Delta x_t} \Delta x_t G_t - \gamma \sigma^2 \Delta x_t\, x_{t-1} - \frac12 \lambda \Delta x_t^2 - \lambda_0 \lvert \Delta x_t \rvert - \frac12 \gamma \sigma^2  \Delta x_t^2 \ .
\end{equation}
Being sloppy with the non-differentiability of the absolute value function, the optimality equation can be written as
\begin{equation}
(\gamma\sigma^2 + \lambda) \Delta x_t = G_t - \gamma \sigma^2 x_{t-1} - \lambda_0 \sgn(\Delta x_t) \ .
\end{equation}
This equation has a non zero solution only if
\begin{equation}
\left\lvert \frac{G_t}{\gamma \sigma^2} -  x_{t-1} \right\rvert > \frac{\lambda_0 }{\gamma \sigma^2}\ ,
\end{equation}
which reminds why spread costs can be thought of as a threshold on the predictive signals \citep[see][for example for a better treatment]{JdL}. The sum of the signals must be large enough for the trade towards Markowitz position to be larger than a size given by the right-hand side of the last equation.

Let us see how this threshold behaviour appears when spread costs are expressed as effective quadratic costs.
The optimal trade verifies the following equation:
\begin{equation}
(\gamma\sigma^2 + \lambda + 2\eta_t) \Delta x_t = G_t - \gamma \sigma^2 x_{t-1} \ .
\end{equation}
If optimal trade is not zero, the constraint $-s \leqslant \Delta x \leqslant s$ is saturated:
\begin{equation}
\frac{\lvert \Delta x \rvert}{s} = 1 \ .
\end{equation}
From \eqref{eq_on_s}, $s = \lambda_0 / (2\eta)$ and the ratio can be written
\begin{equation}
\frac{\lvert \Delta x \rvert}{s} = \frac{2\eta}{\gamma\sigma^2 + \lambda + 2\eta} \frac{\lvert G_t - \gamma \sigma^2 x_{t-1} \rvert}{\lambda_0} \ .
\end{equation}
As a function of $\eta$, the ratio increases from 0 when $\eta = 0$ to $\lvert G_t - \gamma \sigma^2 x_{t-1} \rvert / \lambda_0$ when $\eta \to \infty$. The ratio can cross 1 for a finite $\eta$ if and only if $\lvert G_t - \gamma \sigma^2 x_{t-1} \rvert / \lambda_0 > 1$, which is the threshold condition shown earlier.

To summarise, if the threshold is reached, there exist finite effective quadratic costs that account for the spread costs. Otherwise, effective quadratic costs are infinite and optimal trade is 0.

A similar analysis can be done for a $L^1$ position term. We remind that the constraint on position reads:
\begin{equation}
-u^i_t \leqslant x^i_{t-1} + \Delta x^i_t \leqslant u^i_t \ .
\end{equation}
Equation \eqref{eq_on_s} is replaced by
\begin{equation}
-\lambda_l + 2\eta_i u_i = 0 \ .
\end{equation}
In the one-stock static model case, the equation verified by the optimal trade is
\begin{equation}
(\gamma\sigma^2 + \lambda + 2\eta_t) \Delta x_t = G_t - (\gamma\sigma^2 + 2\eta_t) x_{t-1} \ .
\end{equation}
As before, we compute the ratio that is equal to 1 when the constraint is saturated:
\begin{equation}
\frac{\lvert x_{t-1} + \Delta x_t \rvert}{u_t} = \frac{2\eta_t}{\gamma\sigma^2 + \lambda + 2\eta_t} \frac{\lvert G_t + \lambda x_{t-1}\rvert}{\lambda_l} \ .
\end{equation}
For next-step position $x_t$ to be non zero, we must have
\begin{equation}
\lvert G_t + \lambda x_{t-1}\rvert > \lambda_l \ .
\end{equation}
We remain in position if twice the cost incurred if we cut the position plus the expected returns is greater than the cost to finance the current position. This can be seen by multiplying both sides by $\lvert x_{t-1} \rvert$:
\begin{equation}
\lvert G_t x_{t-1} + \lambda x_{t-1}^2\rvert > \lambda_l \lvert x_{t-1} \rvert \ .
\end{equation}
In the case where $G_t = 0$, if it is cheaper to cut the position then buy it back than to finance it overnight, the optimiser should cut it so that other positions can be taken and financed. The predictive signals modulate the comparison by cutting the position sooner if the prediction is in the opposite direction from the position or by maintaining the position despite financing costs if the position is expected to earn enough.

When we perform signal-wise attribution of these $L^1$ terms, we see that signal-wise trades (resp. positions) are non zero only if the total trade (resp. position) is also non zero. If the sum of the signals does not reach the threshold, no signal gets a trade nor maintain a position. This suggests an interpretation of these terms / constraints as voting systems. If no ``agreement'' is reached between signals, nothing is done.

This attribution along with its interpretation can be profitably used in trading systems where predictive signals are relatively small in front of high spread costs\footnote{or costs induced by taxes such as stamp duties or financial transaction taxes} or in trading system running under tight leverage constraints\footnote{including funds like 130/30 which can be seen as funds with a leverage of 160\% and a net exposure of 100\%}.

\section{Conclusion}

We described a new method that allows to straightforwardly and exactly attribute the effect of constraints to predictive signal portfolios. In all the cases where a distinct portfolio for a constraint or a cost term leads to an awkward interpretation, this attribution allows to cleanly identify the impact of constraints on the signal. From such an attribution a manager is able to make decisions based on the perturbed signal performance for example, or a transfer coefficient can be computed for each signal to assess their implementation in presence of the other signals. We get the closest equivalent of what we would get if we sub-optimally optimised a separate portfolio for each signal under a set of constraints which for each signal would attempt at mimicking the effect of the global constraints. Here, we get the same thing while being optimal, correctly taking into account the constraints and having a perfect split between signals.

Furthermore, as this attribution is totally compatible with the \citet{GrinoldEaston} attribution, we could imagine getting the best of both worlds. For example, let us imagine a trading system where spread costs are high and we have maximum position size constraints that act as safeguards and are thus expected to play little role. It would make sense to see spread costs as effective quadratic costs so that we could attribute them to each signals while identifying a separate portfolio for the constraints in order to monitor their impact and their effectiveness as a whole.

Taking a step back from attribution and considering only the equivalence between constraints and effective risk and costs, the explicit relationship we showed between what we called attribution multipliers and the original Lagrange multipliers generalises the equivalence between bounds and shrinkage as noted by \citep{JMa,Roncallishrink} and let us think of the recent results regarding factor-alignment problems \citep{LeeStefek,BLS,SaxenaStubbs1,SaxenaStubbs2,CSS} as setting up an explicit constraint on an additional factor dependent on the optimal portfolio, which may be easier to handle and more intuitive to understand for a manager than to augment the quadratic risk matrix with the factor projection whose weight is not straightforward to calibrate.

This method can also be seen as a generalisation of the idea of custom risk model\footnote{Note though that the technique called custom risk model also includes a calibration part that is out of the scope of our technique.}. We not only found the natural custom dynamic risk factors associated with constraints, but also effective quadratic costs for constraints or terms in the objective function that are naturally expressed as such, which address some of the concerns expressed in \citep{CSS} regarding the difficulty of finding the correct custom risk factor for a long-only constraint for example. As can be seen from our method, such a factor would indeed vary a lot in time, because at each time step different stocks would be constrained. But it is now possible to compute it explicitly and to try and model it so that an estimate of it be added in the quadratic risk model at the next-step portfolio optimisation.
As we focused here on attribution, we shall not continue in this direction and leave it for future work.

Notwithstanding potential applications in the aforementioned subjects, we would like the reader to consider this signal-wise attribution as an additional item in the toolbox of performance analysis, which yields results that are easy to understand, especially in some cases where other attributions do not and whose interpretation sheds complementary light on how the constraints affect the portfolio and its drivers, the predictive signals.

\section*{Acknowledgements}
We would like to thank Jean-Philippe Bouchaud, Laurent Laloux, Charles-Albert Lehalle and Thierry Roncalli for their comments and fruitful discussions.

\newpage
\appendix

\section{Power-3/2 cost terms}

Power-3/2 cost term can be treated in a similar way as spread costs. For simplicity we shall consider a one-stock static model, but this could be generalised.
The convex optimisation problem
\begin{equation}
\max_{\Delta x_t} \Delta x_t G_t - \lambda_{1/2} \lvert \Delta x_t \rvert^{3/2} - \frac12\gamma \sigma^2 x_t^2
\end{equation}
can be turned into the following equivalent problem, introducing an auxiliary variable $s$ and removing terms independent from $\Delta x_t$:
\begin{equation}
\begin{gathered}
\max_{\Delta x_t, s_t} \Delta x_t G_t - \lambda_{1/2} s_t - \gamma \sigma^2 \Delta x_t\, x_{t-1} - \frac12\gamma \sigma^2 \Delta x_t^2 \\
\lvert \Delta x_t \rvert^{3/2} \leqslant s_t\ .
\end{gathered}
\end{equation}
Note that the constraint is convex as it is the epigraph of a convex function. The constraint is equivalent to the following one
\begin{equation}
\lvert \Delta x_t \rvert \leqslant s_t^{2/3}\ .
\end{equation}
This constraint can be squared and the corresponding augmented objective function is:
\begin{equation}
{\cal F}(\Delta x_t, s_t) = \Delta x_t (G_t - \gamma \sigma^2 x_{t-1}) - \frac12\gamma \sigma^2 \Delta x_t^2 - \lambda_{1/2} s_t - \eta \left( \Delta x_t^2 - s_t^{4/3} \right) \ .
\end{equation}
Optimality conditions corresponds to the critical points of this augmented function:
\begin{align}
(\gamma \sigma^2 + 2\eta) \Delta x_t &= G_t - \gamma \sigma^2 x_{t-1} \label{effcost} \\
\frac43 \eta s_t^{1/3} &= \lambda_{1/2} \ .
\end{align}
Note that equation \eqref{effcost} is familiar as it is the same equation as for quadratic costs. In this case there are only effective quadratic costs as the original quadratic cost term has been replaced by a power-3/2 cost term. We compute the ratio that is equal to 1 when the constraint is saturated:
\begin{equation}
\frac{\lvert \Delta x_t \rvert}{s_t^{2/3}} = \frac{16}9 \frac{\eta^2}{\gamma \sigma^2 + 2\eta} \frac{\lvert G_t - \gamma \sigma^2 x_{t-1} \rvert}{\lambda_{1/2}} \ .
\end{equation}
In this case, the ratio increases from 0 when $\eta = 0$ to $+\infty$ when $\eta \to \infty$ so it always reaches 1. Indeed with these costs there is no threshold effect. It is now straightforward to attribute the total trade to each signal, as explained in the main text.

Note that it is perfectly possible to mix spread costs and power-3/2 costs. This is left as an exercise for the reader.

Last but not least, this method can be generalised to other cost functions $f(\Delta x_t)$ as long as they are convex (for a solution to be found easily by a specialised algorithm) and provided that their reciprocal function is easy to compute, as it is used to get the constraint involving the auxiliary variable $s_t$ under the form
\begin{equation}
\lvert \Delta x_t \rvert \leqslant f^{-1} (s_t)
\end{equation}
where $f^{-1} \circ f = \id$. We shall not pursue this further as a few alternatives to power-3/2 cost term exist (but see \citet{logq} for an example where $f(x) = x \log x$).

\end{document}